# Predicting magnetization of ferromagnetic binary Fe alloys from chemical short range order

**Van-Truong Tran*, Chu-Chun Fu‡ and Kangming Li**

DEN-Service de Recherches de Métallurgie Physique, CEA, Université Paris-Saclay, F-91191 Gif-sur-Yvette, France

Emails: *vantruong.tran@cea.fr and ‡chuchun.fu@cea.fr

## Abstract

Among the ferromagnetic binary alloys, body centered cubic (bcc) Fe-Co is the one showing the highest magnetization. It is known experimentally that ordered Fe-Co structures show a larger magnetization than the random solid solutions with the same Co content. In this work, based on density functional theory (DFT) studies, we aim at a quantitative prediction of this feature, and point out the role of the orbital magnetic moments. Then, we introduce a DFT-based analytical model correlating local magnetic moments and chemical compositions for Co concentrations ranging from 0 to 70 at.%. It is also extended to predict the global magnetization of both ordered and disordered structures at given concentration and chemical short range orders. The latter model is particularly useful for interpreting experimental data. Based on these models, we note that the local magnetic moment of a Fe atom is mainly dictated by the Co concentration in its first two neighbor shells. The detailed local arrangement of the Co atoms has a minor effect. These simple models can fully reproduce the difference in magnetization between the ordered and disordered Fe-Co alloys between 30% and 70% Co, in good agreement with experimental data. Finally, we show that a similar model can be established for another bcc binary Fe alloy, the Fe-Ni, also presenting ferromagnetic interactions between atoms.

## I. Introduction

Binary bcc Fe systems alloyed with a *3d* magnetic element (Ni, Co, Mn, Cr) are known to exhibit distinct host-solute magnetic interaction tendencies, depending on the *d*-band filling of the solute, that is, ferromagnetic (FM) for Ni and Co, with an increased number of *d*-electrons than Fe, and anti-ferromagnetic (AF) for Mn and Cr. Among these alloys, FM Fe-Co exhibits the strongest magnetization.[1] The excellent magnetic properties of Fe-Co alloys have attracted substantial attention from scientists since the pioneering work of Elmen in 1929.[2] It was shown that in the presence of Co solutes, the magnetization of the system is remarkably enhanced, as the magnetic moment of Fe was determined to increase from 2.2 $\mu_B$ to around 3.0 $\mu_B$, depending on the concentration of Co content.[3] The outstanding magnetic properties of Fe-Co alloys make them suitable for the construction of high-quality magnets, which are used for numerous applications, from commercial products to military applications, also for aerospace technologies.[4,5] In addition, recent studies of the magnetic anisotropy in Fe-Co thin films have been carried out for potential applications in the magneto-resistive random access memory devices.[6–9]

The neutron diffraction scattering measurements conducted by Collins and Forsy in 1963[3] have determined the average magnetic moment of each alloy component, showing an enhancement for Fe while the magnetic moment of Co stays almost constant. That study examined the magnetization of bcc Fe-Co during an ordered-disordered transition, but it could not observe a difference of the magnetization between the ordered and the disordered structures. This may be due to the fact that the work investigated the impact of chemical ordering only for two samples at 70% Co. In 1969, Bardos examined more carefully the role of ordering and disordering for Co concentrations ranging from 5% to 70%.[10] The disordered samples were obtained by quenching alloys from high temperatures while the ordered ones were collected by cooling slowly the samples. The measurements revealed that the magnetization of the ordered structures is overall larger than that of the disordered ones. And the difference is most significant around 50 at.% Co. That work,[10] also pointed out that the magnetization of disordered systems reaches a maximum at about 30 at.% Co.

Besides these experimental observations, various theoretical studies have been carried out to investigate thermodynamic and magnetic properties of Fe-Co alloys.[1,6,11–19] For instance, all calculations reported negative mixing enthalpies in Fe-Co systems at all concentrations, indicating a strong ordering tendency.[11,15,17] Indeed, the presence and high stability of the B2 phase is well known both experimentally and theoretically. The enhancement of the magnetic moment of Fe atoms in the presence of Co solutes was well interpreted by examining the electronic behavior by means of DFT calculations,[20] and via a band filling model within the tight binding scheme.[1,12,21] The maximum peak in the magnetization was obtained by previous calculations.[12,15,20] However, the difference in magnetization between the ordered and disordered structures is not fully quantified. Furthermore, a thorough understanding of the interplay between the magnetism and the chemical ordering in Fe-Co alloys is still missing.

The present study is an attempt to quantify the role of chemical ordering/disordering on the magnetization in Fe-Co systems within the ferromagnetic phase. We performed DFT calculations to determine the local and global magnetic properties of Fe-Co alloys for a broad range of Co concentrations (from 0 to 70 at.%) and various degrees of chemical order. The obtained global magnetizations are closely compared with available experimental data. The DFT results are then used to parameterize analytical models for a prediction of local magnetic moments and the global magnetization as functions of local chemical composition and chemical short-range order. Finally, we also construct similar models for another binary FM alloy, the bcc Fe-Ni in order to show the transferability of the present approach.

The paper is organized as follows: a description of chemical ordering/disordering based on the short-range order and the details of DFT calculation setup are presented in Sec. II. The results are discussed in Sec. III, where we first compare the DFT results with experimental data (Sec. III A), and then provide an analysis of electronic structures and local magnetic moments (respectively Sec. III B and III C). Sec. III D and E are devoted to the construction of models correlating magnetism and chemical composition and ordering for respectively Fe-Co and Fe-Ni systems. Finally, conclusions are given in Sec. IV.

## II. Methods

### A. Short-range order parameters of ordered and disordered alloys

In DFT simulations, due to the limited size of supercells considered and the application of periodic boundary conditions, we can only mimic the random solid solutions by quasi-disordered structures, or more precisely the special quasi-random (SQS) structures, with a minimum short-range order. In order to better characterize ordered and quasi-disordered structures, a concept of the chemical short-range order (SRO) are utilized. First, the local arrangement of atoms can be determined by a set of local parameters at each lattice site, which is expressed, for an alloy composed of two species A and B, by the Warren-Cowley formula:[22,23]

$$\alpha_{B_i}^n = 1 - \frac{Z_{A-B_i}^n}{Z^n\left(1 - X_B\right)}, \tag{1}$$

here $\alpha_{B_i}^n$ is the parameter calculated for the n-*th* neighboring shell of a B atom, $Z_{A-B_i}^n$ is the number of A atoms in the n-*th* neighbor shell of the B atom. $Z^n$ is the total number of atoms in the n-*th* shell. In bcc structures, $Z^1 = 8$ and $Z^2 = 6$ etc.. $X_B$ is the global concentration of the B atoms in the alloy. The SRO parameter (on the B atoms) is defined as the average of the individual $\alpha_{B_i}^n$ over all the B atoms:

$$\left\langle \alpha_B^n \right\rangle = \frac{\sum_{i=1}^{N_B} \alpha_{B_i}^n}{N_B} \tag{2}$$

In Fig. 1, structures with various SROs are sketched. For the B2 structure, each Fe (resp. Co) atom is surrounded by 8 and 0 Co (resp. Fe) neighbor atoms in the first and second shells, respectively. It has, therefore, a SRO parameter equal to -1 for the first shell and 1 for the second shell, denoted as [-1, 1].

The number of the nearest neighbor shells of interest can be determined from an evaluation of the relevance of each shell to some most important properties of the alloys, for instance the pairwise binding energies. The binding energy of a A-A pair in the B-atom system is determined as follows:

$$E^{\text{bind}}\left(\text{A-A}\right) = 2E^{\text{tot}}\left((n-1)\text{B}+1\text{A}\right) - E^{\text{tot}}\left((n-2)\text{B}+2\text{A}\right) - E^{\text{tot}}\left(n\text{B}\right) \tag{3}$$

The binding energies of Co-Co (resp. Fe-Fe) pairs in a bcc Fe (resp. Co) lattice were calculated via DFT up to the 4$^{\text{th}}$ neighbor shell and shown in Fig. 2. Only the first two shells are noted as relevant, since the energy magnitudes beyond are smaller than 0.01 eV, and considered as negligible.

The SRO parameters of the cI16-$Fe_7Co$ in Fig. 1(b) and the D0$_3$-$Fe_3Co$ in Fig. 1(c) are [-0.14286, -0.14286] and [-0.33333, -0.33333], respectively. For quasi-disordered structures, the SRO parameters can be written as $\left\langle \alpha_B^n \right\rangle = 1 - \frac{\left\langle Z_{A-B}^n \right\rangle}{Z^n\left(1 - X_B\right)}$, with $\left\langle Z_{A-B}^n \right\rangle = \sum_{i=1}^{N_B} Z_{A-B_i}^n / N_B$ being the average number of atoms A in the n-*th* shell of atoms B, which is obviously equal to $Z^n X_A = Z^n\left(1 - X_B\right)$ due to the possibility of occupation in completely disordered structures. Thus, the disordered alloys have $\left\langle \alpha_B^n \right\rangle = 1-1 = 0$ for all shells.

## B. Details of DFT simulations

The first principles calculations were carried out within the Kohn-Sham density functional theory[24,25] with plane-wave basic sets and projector augmented wave (PAW) potentials as implemented in the Vienna Ab-initio Simulation Package (VASP). [26–29] The calculations were done with the GGA-PBE functional which is widely used in solid state studies, in particular for Fe-alloys.[30] Spin polarization within the collinear approximation was chosen. In all calculations, energies were converged with the criterion of $10^{-6}$ eV in electronic self-consistent loops, and studied structures were fully relaxed until the forces on atoms are less than 0.01 eV/ Å and stresses on the cells are less than 5 kbar. To have a good convergence of energetic, magnetic and structural properties, the plane-wave energy cut-off is chosen equal to 400 eV. The Methfessel-Paxton smearing function with a width of 0.1 eV was used. The Brillouin zones were sampled by the Monkhorst-Pack method[31] with a k-mesh of 16×16×16 for the simple cubic unit cell or equivalent for a supercell: for example, a k-mesh grid of 4×4×4 was used for a supercell of 4×4×4 containing 128 atoms. In this study, all the quasi-random (SQS) and the partially ordered structures of Fe-Co and Fe-Ni are represented by the 128-atom supercells, while the unit cells used for the ordered phases are case dependent.

# III. Results and discussions

## A. Magnetization of ordered and disordered Fe-Co structures

Experimental data of the mean magnetic moment in disordered and ordered Fe-Co structures were obtained by Bardos in 1969.[10] The results show that the magnetization of quenched (disordered) structures increases with the increase of Co content with the maximum of 2.45 $\mu_B$ around 30 at.% Co. On the other hand, the value obtained in ordered structure is almost identical to that of the disordered structures below 30 at.% Co, but the former is significantly higher than the latter in the range of Co concentration from 30% to 70%.

It is worth mentioning that in this paper, the atomic percentage is used to denote the composition of an alloy and thus, for the sake of simplicity, the prefix “at.” is not explicitly written hereafter.

To reproduce the results of the disordered systems, we adopt the special quasi-random structures (SQS) which are good representations of a set of random alloys.[32–34] In order to capture the properties of the ordered structures in this experiment, we replaced randomly Fe atoms by Co ones in a bcc Fe lattice up to a given Co concentration and then searched for configurations having the SRO parameters as close as possible to the ones of the stoichiometric B2 phase. These structures may be called as B2-like structures. For example, in the supercells of 128 atoms, the closest SRO parameters are [-0.88235, 0.91634] and [-0.54217, 0.70299] at 46.88% and 35.16% Co, respectively. We have checked that a B2-like structure presents a lower energy than the SQS structure with the same Co content. As shown in the phase diagram of Fe-Co, low temperature structures transform from bcc to hcp (hexagonal close-packed) above 70% Co.[16,35] We therefore limit this study to Fe-Co alloys with Co concentration less than 70%, as bcc is the phase of our interest.

As expected, our DFT calculations found the ferromagnetic (FM) state to be the ground state of all bcc Fe-Co structures. To verify the DFT data, the experimental magnetization values are also extracted from Ref. 10 for a comparison and all the data are shown in Fig. 3.

First, we notice a qualitative agreement between DFT and experiments on the composition dependence of the magnetization. But, a clear quantitative discrepancy is present, as can be seen in Fig. 3(a). Some factors may contribute to such disagreement. For instance, experimentally, the magnetic moments at 0 K were not measured directly but extrapolated from values obtained at 295 K. Details of the experimental set-up and the extrapolation are described in Refs.10 and 36. Also, the magnetic moment obtained from experiments is actually the sum of the spin and orbital moments, while in our DFT calculations, only the spin moment contributions were computed. For Fe-Co bulk systems, it was demonstrated that the spin magnetic moment is the main part of the total magnetization, and the orbital magnetic moment plays a minor role with a contribution of about 4% in pure Fe and 8% in pure Co.[37] In principle, the orbital moments can be determined from DFT by systematically including the spin-orbit coupling in calculations for all the alloy concentrations. But it is too CPU demanding. Alternatively, the orbital moments can be estimated from the spin moment (or the total moment) in the Kittel's and Van Vleck's theory via a splitting factor[38,39]

$$g = 2\left(1 + \frac{M_{orb}}{M_{sp}}\right) \quad (4)$$

For systems where $M_{orb}$ is negligible, $g \approx 2$ and this coefficient is larger than 2 if the orbital contribution is taken into account. By using Eq. (4) and the relationship $M_{tot} = M_{orb} + M_{sp}$, we have

$$\begin{cases} M_{orb} = \dfrac{g-2}{g} M_{tot} \\ M_{sp} = \dfrac{2}{g} M_{tot} \end{cases} \quad (5)$$

Thus, by knowing the splitting factor *g*, we can separate the contributions from spin and orbital moments to the total magnetic moment. In practice, *g* can be determined via the factor *g′*, which can be measured in magnetomachanical experiments, via the following equation[40]

$$\frac{1}{g} + \frac{1}{g'} = 1 \quad (6)$$

In the work of Mayer[40], $g'$ was experimentally determined to be 1.928 and 1.854 for Fe and Co, respectively. This leads to $g_{Fe}$ = 2.076 and $g_{Co}$ = 2.170.

For Fe-Co alloys, an estimation of the effective *g* factor can be calculated via the following equation[40]

$$g_{\mathrm{eff}} = \frac{M_{Fe} \times X_{Fe} + M_{Co} \times X_{Co}}{M_{Fe} \times X_{Fe} / g_{Fe} + M_{Co} \times X_{Co} / g_{Co}}, \quad (7)$$

in this equation *M* is the total magnetic moment. The effective *g* factor is plotted as an inset in Fig. 3(b), where the black curves are the spin magnetic moment estimated from the experimental data by using Eqs. (5) and (7). It can now be seen that the DFT results are in better agreement with the experimental ones, particularly on the right side of the peak. The remaining visible discrepancy on the Fe-rich side could come from the fact that the Fe moment in Fe-Co can only saturate at around 2.7 $\mu_B$ in DFT calculations as raised in previous studies with both LSDA[19] (local spin-density approximation) and GGA[41] functionals. And it is below the value of 3.0 $\mu_B$ observed in the experiment.[3] Our DFT data also reveal a underestimation of the saturated Fe moment, i.e. 2.76 $\mu_B$ (the highest magnetic moment obtained in the Fe-Co B2 structure) compared to 3.0 $\mu_B$ from experiments.[3]

In spite of the remaining discrepancy, with an estimation of the orbital moment, our DFT results are closer to the experimental data compared to the outcome from previous calculations by Díaz-Ortiz et al.,[15] and show clearly that ordered structures with a relatively high Co concentration exhibit a higher moment than disordered ones. Although the peak in the DFT curve of the disordered structures is observed at about 23% Co, smaller than 27% in the experiment, both show a maximum magnetic moment of around $\langle M \rangle$ = 2.34 $\mu_B$. In addition, we want to point out that, the enhancement of the magnetization of the ordered structure compared to that of the disordered ones is very well reproduced by our DFT data as shown in Fig. 3(c). The largest discrepancy is observed at 50% Co, where the difference between ordered and disordered magnetizations from DFT and experiments are 0.079 $\mu_B$ and 0.068 $\mu_B$, respectively.

To further analyze the magnetization, we separate the contribution of individual Fe and Co atoms. The average moment of each species is shown in Fig. 4. In both ordered and disordered structures, the magnetic moment of Co just varies slightly with the change of Co content. This result is consistent with experimental observations.[3] On the other hand, the moment of Fe is strongly affected by the presence of Co solutes. In the SQSs, the average moment of Fe atoms increases rapidly from around 2.2 $\mu_B$ to 2.5 $\mu_B$ in the range from 0 to 20% Co, then it varies slowly up to a saturated value of 2.6 $\mu_B$ and remains almost constant above 50% Co. In the ordered structures, the magnetic moment of Fe increases with Co concentration and reaches a maximum at 50% Co before decreasing. The position of this maximum is consistent with the fact that the ordered 50% Co structure, i.e. the stoichiometric B2 phase, is the most stable phase of Fe-Co alloys at low temperatures and possesses the highest magnetization.[12]

### B. Electronic structures of Fe and Co components

To better understand the magnetic behavior in Fig. 4, we investigate the electronic properties in Fe-Co structures and compare them to those in pure Fe and Co ones.

In Fig. 5, the left panels of both Fig. 5(a) and Fig. 5(b) present the electronic band structures along the high symmetry paths in pure FM bcc Fe and Co structures. In all the panels, the dot lines are the spin down contribution while the solid ones show the spin up results. The right panels are the local density of states (DOS) of Fe and Co atoms. The black lines are the DOS of these atoms in their pure systems and the color (red and blue) curves are the local DOS in the Fe-Co B2 structure.

First, from the band structure in Fig. 5(a) we see that just above the Fermi level, the unoccupied bands are mostly spin down ones. On the right panel, the sharp peaks in both the spin down and spin up DOS are originated from the nearly flat bands in the left panel. The same behavior also holds for the energy bands and the DOS in the case of Co shown in Fig. 5(b). Interestingly, the red lines show that both the spin up and spin down DOS of Fe atoms in Fe-Co B2 are shifted down compared to the black ones in pure Fe, this moves the spin-up peaks far away from the Fermi level while the one of spin down goes closer to the Fermi level. This enhances charge density and thus the magnetic moment of Fe atoms, which reflects the results shown in Fig. 4.

For the pure bcc Co, the bands and the DOS look quite similar to those of pure bcc Fe by just shifting down the whole band structure with respect to the Femi level. As a result, the DOS in pure Co are also similar to that in pure Fe. The DOS of Co atoms in the Fe-Co B2 structure change only slightly compared to the one in pure Co. Therefore, the magnetic moments of Co atoms remain almost the same.

It is also worth noting that the local DOS of Fe and Co in the B2 structure are almost the same as can be seen from the inset in the DOS panel in Fig. 5(b). We can thus conclude that Fe atoms in the B2 structure reach their magnetically saturated state due to the interaction with Co atoms.

## C. Dependence of magnetization on local chemical environment

All the above results only reveal the dependence of the magnetization on the global concentration of Co. To understand the impact of local chemical compositions on the magnetic behavior, we analyze the DFT data from all the considered SQS and ordered structures. The resulting distribution of Fe and Co local moments are shown in Fig. 6.

It can be seen that in the ordered structures, most Fe atoms have only Co atoms as first neighbors, and a high magnetic moment of around 2.7 $\mu_B$ while Fe moments in the SQS structures vary from around 2.2 $\mu_B$ to 2.6 $\mu_B$. All the magnetic moments of Co atoms are mainly concentrated in the 1.7 $\mu_B$ peak. This is consistent with the data given in Fig. 4, and the fact that Co is magnetically saturated, showing a fully occupied majority spin band (Fig. 5b).

To understand the strong variation of the Fe moments, particularly in the SQS structures, we analyze the magnetic moment of each Fe atom in all the configurations with respect to the number of Co neighbors in the nearest and next-nearest shells. All the data are depicted in Fig. 7(a).

The atoms with the same color have the same number of Co neighbor atoms $N_1$ in the first shell. The projections along $N_1$ and $N_2$ are also shown in Fig. 7(b) and Fig. 7(c), respectively. As can be observed, the moment on each Fe atom depends strongly on the number of Co nearest neighbors $N_1$. The Fe moment increases remarkably with the increase of $N_1$, from 2.2 $\mu_B$ to about 2.8 $\mu_B$. The dispersion of the Fe moments at a given $N_1$ in Fig. 7(b) is due to different Co atom occupation at the second shell $N_2$ and, in a minor extent, at farther neighboring shells. With the same number of Co atoms $N_1$ at the first shell, the additional presence of $N_2$ Co neighbors appears to have a weaker impact, although still visible (Fig. 7(c)) and needs to be considered. It is also worth noting that the filling effect of the second shell depends on the occupancy of Co at the first shell. When Co atoms occupy 50% of the first-shell sites, the Fe moment is less sensitive to $N_2$. However, if there is no Co atom in the first shell (black points in Fig. 7), the filling of Co at the second shell significantly enhances the moment on the center Fe atom. Finally, when the first shell is fully occupied by Co atoms (violet points), the additional filling of Co atoms at the second shell tends to slightly decrease the Fe magnetic moment.

To verify if distinct arrangements of the same number of Co neighbors could significantly affect the Fe moment we examined three different configurations with the same number of Co atoms (3 Co atoms) but located at different positions at the first shell of a Fe atom. These three configurations and the local DOS of the central Fe atom are shown in Fig. 8. As can be noticed, the local DOS and the magnetic moment of the Fe atom are almost identical for the three cases, suggesting that this effect is negligible.

## D. Simple analytical models to predict magnetism in Fe-Co alloys

### *1. An atomic model for Fe local magnetic moments*

The above analysis reveals that the magnetic moment of a Fe atom depends strongly on the local environment characterized by the number of Co neighbor atoms. Although the first shell is the dominant one to be considered, the impact of the second shell is also relevant. Therefore, to capture the variation of Fe moments, we introduce an analytical expression that containing both $N_1$, $N_2$ as follows:

$$M_{Fe} = P_0 + A_1 \times N_1 + A_2 \times N_1^{\ 2} + A_3 \times N_1^{\ 3} + B_1 \times N_2 + B_2 \times N_2^{\ 2} + B_3 \times N_2^{\ 3} \\ + C \times N_1 N_2 + D_1 \times N_1^{\ 2} N_2 + D_2 \times N_1 N_2^{\ 2} \qquad (8)$$

The introduction of high-order terms in $N_1$ and $N_2$ in Eq. (8) could be useful to describe the curved behavior around 25% Co and the moment saturation around 50% as shown in Fig. 4. The cross terms, which contain both $N_1$ and $N_2$, are used to capture the correlated effects of the first and second shells. To determine all the parameters $P_0$, $A_1$, $A_2$, $A_3$, $B_1$, $B_2$, $B_3$, $C$, $D_1$ and $D_2$, we employed a least square fitting approach in which values of $M_{Fe}$ and $N_1$, $N_1^2$, $N_1^3$, $N_2$, $N_2^2$, $N_2^3$, and $N_1N_2$, $N_1^2N_2$, and $N_1N_2^2$ of each Fe atom came from our DFT calculations. The DFT data of each Fe atom allows us to construct one equation and a set of about 2200 equations obtained from 20 SQSs and 10 B2-like structures was set up for fitting. The obtained parameters are listed in Table I. A simpler version of Eq. (8) with fitting only up to the second order in $N_1$, $N_2$ was also performed.

As can be observed from Table I, the fitted parameters $A_1$, $A_2$, $A_3$, $B_1$, $B_2$, and $B_3$ indicates that the impact of the first shell is almost twice as strong as that of the second shell.

To check the accuracy and usefulness of this analytical model, it was applied to calculate the magnetic moment of all the Fe atoms in each Fe-Co system based on the respective $N_1$, $N_2$. The predicted atomic moments from the model were then used to calculate the Fe average moments and the magnetization of the structures, where the magnetic moment of each Co atom is considered to be 1.76 $\mu_B$ for all concentrations.

The results obtained from this atomic analytical model are plotted in Fig. 9. From Fig. 9(a), the model reproduces very well the magnetization in both the SQSs and the ordered structures with all important features pointed out by DFT and experiments in Fig. 3. The model also correctly predicts magnetic moment variation of Fe with increasing Co content for both the SQSs and the B2-like ordered structures, as shown in Figs. 9(b) and (c), respectively.

Then, the model is further verified in calculating moments in additional structures which were not included in the fitting database. We employed the analytical formula (8) to calculate the average magnetic moment of Fe in various well-known ordered structures B2, B32, cI16, and $D0_3$.[12] A comparison between DFT and the model fitted up to the third order in $N_1$, $N_2$ is shown in Fig. 10, and the values are given in table II. The results obtained by the model on fitting just up to the second order were also calculated for a comparison. Fig. 10 shows a good agreement between our model and DFT results for all the considered structures. Please note that the visible discrepancy in $Co_7Fe$ is because this structure has 87.5% Co and Co atoms fill completely the 1st and 2nd shells of some Fe atom. Such configurations were not represented in the database of fitting that limited to 70% of Co as we explained above.

### *2. A SRO-based analytical model for average magnetic moments*

The above model (Eq. 8) enables to predict local Fe moments from the detailed information of the atomic structures. When such information is inaccessible, for instance in most experiments, it is useful to extend the atomic model to estimate the average magnetic moment of Fe and thus the magnetization through global Co concentration and chemical SRO parameters which can be determined in for example neutron scattering experiments.[42–44] From Eq. (8), the average moment of Fe can be deduced as

$$\begin{aligned}\langle M_{Fe}\rangle = \frac{\sum_{i=1}^{N_{Fe}} M_{Fe_i}}{N_{Fe}} &= P_0 + A_1\times\langle N_1\rangle + A_2\times\langle N_1^2\rangle + A_3\times\langle N_1^3\rangle \\ &+B_1\times\langle N_2\rangle + B_2\times\langle N_2^2\rangle + B_3\times\langle N_2^3\rangle + C\times\langle N_1N_2\rangle + D_1\times\langle N_1^2N_2\rangle + D_2\times\langle N_1N_2^2\rangle\end{aligned}, \quad (9)$$

here $\langle N_1\rangle$ and $\langle N_2\rangle$ are the average number of Co neighbor atoms in the 1st and 2nd shells of a Fe atom. Strictly speaking, $\langle N_1^2\rangle$, $\langle N_2^2\rangle$ and $\langle N_1N_2\rangle$ are generally different from $\langle N_1\rangle^2$, $\langle N_2\rangle^2$ and $\langle N_1\rangle\langle N_2\rangle$ due to the varying numbers of Co neighbors among Fe atoms. Here we consider such difference to be negligible, with an assumption that there is no large composition fluctuation in different regions of an alloy. Thus, we have

$$\langle M_{Fe} \rangle \approx P_0 + A_1 \times \langle N_1 \rangle + A_2 \times \langle N_1 \rangle^2 + A_3 \times \langle N_1 \rangle^3 + B_1 \times \langle N_2 \rangle + B_2 \times \langle N_2 \rangle^2$$
$$+B_3 \times \langle N_2 \rangle^3 + C \times \langle N_1 \rangle \langle N_2 \rangle + D_1 \times \langle N_1 \rangle^2 \langle N_2 \rangle + D_2 \times \langle N_1 \rangle \langle N_2 \rangle^2 \tag{10}$$

On the other hand, the coefficient $\alpha$ of each Fe atom from Eq. (1) can be determined as $\alpha_{Fe_i}^n = 1 - \frac{Z_{Co-Fe_i}^n}{Z^n (1 - X_{Fe})}$. And the SRO of Fe atoms at the n-*th* shell is defined as the mean value of these coefficients

$$\langle \alpha_{Fe}^n \rangle = \frac{\sum_{i=1}^{N_{Fe}} \alpha_{Fe_i}^n}{N_{Fe}} = 1 - \frac{\frac{\sum_{i=1}^{N_{Fe}} Z_{Co-Fe_i}^n}{N_{Fe}}}{Z^n (1 - X_{Fe})} = 1 - \frac{\langle Z_{Co-Fe}^n \rangle}{Z^n (1 - X_{Fe})}, \tag{11}$$

where $\langle Z_{Co-Fe}^n \rangle$ is the average number of Co atoms at the n-*th* shell of a Fe atom, which is actually $\langle N_n \rangle$. Thus we have $\langle \alpha_{Fe}^n \rangle = 1 - \frac{\langle N_n \rangle}{Z^n (1 - X_{Fe})}$ and $\langle N_n \rangle$ can be calculated via the SRO $\langle \alpha_{Fe}^n \rangle$ by equation:

$$\langle N_n \rangle = \left[ 1 - \langle \alpha_{Fe}^n \rangle \right] Z^n X_{Co}, \tag{12}$$

here it has already taken $X_{Co} = 1 - X_{Fe}$. Eq. (12) can be expressed more explicitly for each shell as:

$$\begin{cases} \langle N_1 \rangle = \left[ 1 - \langle \alpha_{Fe}^1 \rangle \right] 8 X_{Co} \\ \langle N_2 \rangle = \left[ 1 - \langle \alpha_{Fe}^2 \rangle \right] 6 X_{Co} \end{cases} \tag{13}$$

Thus, even though only the SRO parameters and Co concentrations are known, we can still use the SRO rule, i.e. Eqs. (13) and (10), to evaluate the average magnetic moment of Fe and thus the magnetization. These two equations establish a SRO-based model.

Considering two extreme cases:

(i) The B2 phase: $\left[\langle\alpha_{Fe}^{1}\rangle,\langle\alpha_{Fe}^{2}\rangle\right]=[\text{-}1,1]$ and thus $\langle N_1\rangle=8$ and $\langle N_2\rangle=0$, Eq. (10) becomes $\langle M_{Fe}\rangle \approx P_0 + A_1\times 8 + A_2\times 8^2 + A_3\times 8^3 = 2.762\ \mu_B$, which agrees well with results using Eq. (8) and also by DFT (see Table II).

(ii) The ideal SQSs should have $\left[\langle\alpha_{Fe}^{1}\rangle,\langle\alpha_{Fe}^{2}\rangle\right]=[0,0]$ and thus $\langle N_1\rangle=8X_{Co}$ and $\langle N_2\rangle=6X_{Co}$.

Eq. (10) for SQSs can be thus written as:

$$\begin{aligned}\langle M_{Fe}\rangle &\approx P_0+\left[8A_1+6B_1\right]X_{Co}+\left[8^2A_2+6^2B_2+8\times 6C\right]{X_{Co}}^2\\&+\left[8^3A_3+6^3B_3+8^2\times 6D_1+8\times 6^2D_2\right]{X_{Co}}^3\end{aligned} \tag{14}$$

This equation is plotted as the solid black line in Fig. 9(b) which proves the validity of the model.

The results for ordered structures predicted using the SRO-based model are compared with those from the atomic model and the DFT calculations in Table II.

We note that the second order models (both atomic and SRO-based) already reproduce well the DFT results of the structures given in Table II and S1. However, this can be problematic for the disordered structures. Indeed, as can be seen from Eq. (14), the second order model leads to a parabolic variation of the average magnetic moment of Fe, and thus fails to capture the magnetic saturation above 50% Co as predicted by DFT in Fig. 4.

To further check the merit of Eqs. (10) and (13), we generated different SRO structures and DFT calculations were carried out to get the magnetic moments of Fe atoms for checking the prediction by the analytical model. Structures with the number of Co atoms equal to 30, 64 and 90 in supercells of 128 atoms representing for low, intermediate and high Co concentrations in Fe-Co systems were examined. And for each concentration, five different SRO configurations were considered. The data calculated by DFT and the analytical model are shown in Table S1 in the Supplemental Material. It can be seen that results using Eqs. (10) and (13) or Eq. (8) are in good agreement with the DFT ones. In both Tables II and S1, the data reveal that the average magnetic moment of Fe calculated by Eqs. (10) and (13) is slightly better than that using Eq. (8) in many cases.

### *3. Explaining the difference in magnetization between ordered and disordered structures*

The enhanced magnetization in ordered structures compared to SQSs can be understood more precisely by looking at the average number of Co neighbor atoms around a Fe atom in these structures.

For SQSs, the SRO of each shell is close to zero, whereas the ordered B2-like systems favor the SRO close to -1 and 1 for the 1st and 2nd shells, respectively. Thus we can write

$$\begin{cases} \left[ \langle \alpha_{Fe}^{1} \rangle_{\mathrm{Ord}}, \langle \alpha_{Fe}^{2} \rangle_{\mathrm{Ord}} \right] = [-1+\eta, 1-\eta] \\ \left[ \langle \alpha_{Fe}^{1} \rangle_{\mathrm{SQS}}, \langle \alpha_{Fe}^{2} \rangle_{\mathrm{SQS}} \right] = [\delta, \delta] \end{cases}, \tag{15}$$

with positive numbers $\eta$ and $\delta$ as close as possible to zero. From Eq. (13), it can be seen that

$$\begin{cases} \langle N_1 \rangle_{\mathrm{Ord}} = [2-\eta] 8 X_{Co}, \langle N_2 \rangle_{\mathrm{Ord}} = \eta 6 X_{Co} \\ \langle N_1 \rangle_{\mathrm{SQS}} = [1-\delta] 8 X_{Co}, \langle N_2 \rangle_{\mathrm{SQS}} = [1-\delta] 6 X_{Co} \end{cases} \tag{16}$$

Thus the ordered structures tend to have more Co atoms in the 1st shell of each Fe atom than that in SQSs, and the situation is reversed in the 2nd shell. However, the impact of the 1st shell is twice as strong as that of the second shell as indicated by the fitting parameters, therefore, the average Fe moment in ordered structures is overall larger than in SQSs.

In practice, we can always obtain a random structure (SQS) at any concentration with $\delta \to 0$ ($\delta$ < $10^{-2}$). However, if searching for B2-like ordered structures, $\eta$ can be very close to zero only for the structures with a concentration close to 50% Co, then this value becomes larger, even reaching 1 when the ordered structure has a concentration far away from 50%. The variation of $\eta$ indicates that the difference of the number of Co neighbor atoms in ordered and disordered structures at the same concentration of Co will be most significant at 50% Co (as $\eta \approx 0$), and it becomes smaller and even negligible (as $\eta \to 1$) for the structures far from 50% Co. Consequently, the difference between the magnetization of ordered and disordered structures is the most significant at 50% and gradually decreases towards the two sides of this concentration. This elucidates the behavior shown experimentally and predicted by calculations, as depicted in Fig. 3.

### *4. SROs for a desired magnetization in bcc Fe-Co*

The set of Eqs. (10) and (13) may be used not only to estimate the magnetization of a given Fe-Co alloy and compare to experimental data but also to predict structures characterized by certain chemical SROs that present high or some desired magnetization, sufficiently below the Curie temperatures.

For the latter, we show in Fig. 11 the Fe moments as functions of the SROs $\langle \alpha_{Fe}^{1} \rangle$ and $\langle \alpha_{Fe}^{2} \rangle$ for three distinct concentrations of Co. In all cases, very high magnetic moments can be obtained in the structures with SRO parameters close to those of the B2 structure ([-1, 1]). In Fig. 11 (a) with 25% of Co, the spectrum indicates that the average moment of Fe in the SQSs is close to the one of the accessible ordered B2-like structure. This could be understood as the dominant SRO parameter of the latter, the 1$^{st}$ shell one (-0.333), is closer to the one of the SQSs than that of the stoichiometric B2 structure, due to the rather low Co concentration. This result also agrees with the experimental evidence that the difference between the magnetization of ordered and disordered structures is negligible below 30% Co.[3]

Figs. 11(b) and 11(c) predict the highest moment in structures with SRO = [-1, -1] which, in the case of $X_{Co} = 0.5$, means that the first and second shells of each Fe atom are completely filled by Co atoms. However, such configuration cannot exist in reality except in the very Co-rich region of the alloy, where the ground state is rather the hcp phase. This region is beyond the scope of the present model. Therefore, the B2-like structures remain the highest magnetic system among the accessible structures. The black region in Fig. 11(c) is the forbidden region due to $\langle N_1 \rangle$ and/or $\langle N_2 \rangle$ being larger than $Z_1 = 8$ and/or $Z_2 = 6$.

In summary, such spectra provide information about the required range of SRO parameters of the alloy at a given concentration, in order to achieve a desired average Fe moment and the global magnetization.

### E. An extension of the analytical models for other alloys: the case of bcc Fe-Ni

As both Co and Ni are close neighbors of Fe in the Periodic Table of Elements, located on the right side in the same 3*d* period, the magnetic properties of bcc Fe-Ni and Fe-Co alloys exhibit similarities, showing an enhanced Fe moment due to the presence of the alloying atoms (Ni or Co). Also, the magnetization of bcc Fe-Ni alloys increases with increasing Ni content up to a maximum at 10% as shown experimentally.[45] In this section, the approach developed in Sec. III D is adapted for the bcc Fe-Ni systems as an example of its transferability to similar systems.

According to the Fe-Ni phase diagram by Cacciamani et al., [46] the bcc phase of Fe-Ni can be retained up to 7% Ni at about 450 °C. However, due to the hysteresis on heating and cooling through the mixed phase region in such alloys, either bcc or fcc (face-centered cubic) phase can be retained under metastable conditions.[45] The experiment conducted by Crangle and Hallam[45] shows that the bcc phase of Fe-Ni can be obtained with the composition ranges up to around 32% Ni. On the other hand, our DFT calculations showed that the bcc quasi-disordered (SQS) structures containing more than 40% Ni become unstable and spontaneously relax towards the body-centered tetragonal (bct) phase. Therefore, only the disordered structures with a concentration less or equal than 40% Ni are considered in this study. Although no ordered structures were found experimentally, two FM ordered structures, cI16-$Fe_7Ni$ and $D0_3$-$Fe_3Ni$ are also included in the fitting data, to account for the effects of chemical ordering on the magnetic moments.

To describe the dependence of the atomic moments on the local environment in the Fe-Ni systems, we employ an analytical model similar to Eq. (8). However, to be able to reproduce the maximum magnetization shown by our DFT calculations and the experiment,[45] our magnetic models for Fe and Ni atoms need to be fitted up to the 3$^{rd}$ neighbor shells and up to a 3$^{rd}$ order polynomial, which can be generalized from Eq. (8) as:

$$M_{ele} = P_0^{ele} + \sum_{i=1}^{3} A_i^{ele} \times N_i + \sum_{i=1}^{3}\sum_{j=i}^{3} B_{ij}^{ele} \times N_i N_j + \sum_{i=1}^{3}\sum_{j=i}^{3}\sum_{k=j}^{3} C_{ijk}^{ele} \times N_i N_j N_k \ , \qquad (17)$$

where $N_i$ is the number of Ni neighbors at the i-*th* shell of a Fe or Ni atom. $P_0^{ele}$, $A_i^{ele}$, $B_{ij}^{ele}$ and $C_{ijk}^{ele}$ are the fitting parameters; *ele* being Fe or Ni.

The model fitting was performed based on DFT data on eight 128-atom SQS supercells of different concentrations and the two ordered systems. It should be pointed out that since we have more fitting data from the SQSs than the ordered structures, the model may be more accurate for the prediction of the magnetic properties of structures with small chemical short-range orders. The fitting parameters for Fe and Ni are given in Table III. Then, the SRO-based analytical model can be derived from Eq. (17) by replacing $N_i$ by the average number $\langle N_i \rangle$, with $\langle N_1 \rangle$, $\langle N_2 \rangle$ calculated using Eq. (13) and $\langle N_3 \rangle = \left[1 - \langle \alpha^3_{Fe/Ni} \rangle\right] 12 X_{Ni}$.

Magnetic moments from our DFT calculations and the atomic and SRO-based analytical models for the SQSs and the ordered Fe-Ni structures are presented in Fig. 12 and Fig. 13, respectively. In Fig. 12, the experimental data on solid solutions from Crangle and Hallam [45] are also shown.

It is worth mentioning that we neglected the impact of the orbital moments in this case as shown in Ref. 40, the effect of the orbital moment in Fe-rich Fe-Ni alloys (here below 40% Ni) is as weak as in pure Fe.

First of all, our DFT results are in good agreement with the experimental values, and successfully predict the maximum at about 10% Ni. The local and SRO-based models also reproduce such a feature in disordered alloys, and accurately describe the magnetization of the ordered structures (compared to DFT). It is noted that similar to the case of Fe-Co, the magnetization of ordered structures is also higher than that of disordered structures in Fe-Ni systems.

The average moments of Fe and Ni are shown in Fig. 13(a) and Fig. 13(b), respectively. Overall, we see that the results from the analytical models (open symbols) present the same behavior as that obtained by DFT (black filled squares), and the models seem to be more accurate for Ni than for Fe moments. Similar to the Fe-Co case, the average magnetic moment of Fe atoms increases with the increase of Ni content (Fig. 4). On the other side, the average moment of Ni atoms shows an opposite trend. As a result, in Fe-Ni alloys, both the magnetic moments of Fe and Ni atoms are enhanced compared to the values in the respective pure systems.

It is worth mentioning that, due to the reduction of the Ni moments versus Ni content, and the smaller Ni moment magnitudes (equal or less than 0.86 $\mu_B$) compared to the constant moment of Co atoms ($\approx 1.76$ $\mu_B$) in Fe-Co, bcc Fe-Ni alloys present smaller magnetizations than bcc Fe-Co as pointed out previously.[5,45]

As for Fe-Co, we validate the analytical models for Fe-Ni by considering some additional Fe-Ni structures not included in the fitting database. These structures are partially ordered with SRO parameters larger than those of SQSs. Results obtained for these structures by both DFT and the models are represented in Table S2 in the Supplemental Material, which once again demonstrates the ability of these analytical models to predict magnetic moments in excellent agreement with DFT data.

## IV. Conclusion

This study aims at determining the role of chemical ordering/disordering on the magnetic properties of ferromagnetic binary Fe alloys for a broad range of concentrations. We first focused on the case of a strong ferromagnet, the bcc Fe-Co, and then transferred the same methodology to the case of bcc Fe-Ni.

First of all, systematic DFT calculations were performed to determine local magnetic moments and global magnetization of these systems with several concentrations (from 0 to 70% Co in Fe-Co, and from 0 to 40% Ni in Fe-Ni) and various degrees of chemical order. The Fe local moments in both alloys tend to increase with an increase of local composition of Co and Ni in respectively Fe-Co and Fe-Ni alloys, while the Co local moment remains constant in all the cases, and the Ni local moment in the considered Fe-Ni alloys decreases with increasing local Ni concentration. Concerning the global magnetization versus solute concentration behaviors, it is worth mentioning that DFT results are in excellent agreement with experimental data for the Fe-Ni alloys, whereas for the Fe-Co, the contribution of the orbital moments (which is more important for Co atoms) has to be removed from the experimental data, in order to achieve a better experimental-DFT agreement. But, the experimental concentration for the highest magnetization is still underestimated by DFT. This fact may be due to an underestimation of the value of the saturated Fe moment within DFT-GGA (2.76 $\mu_B$ instead of 3.0 $\mu_B$), and also raised in previous DFT-LDA studies.

Based on the density functional theory results, rather simple analytical models were constructed for the prediction of local magnetic moments as functions of local chemical compositions in the alloys. The satisfactory accuracy of these atomic models, attested by comparisons with DFT and experimental data, suggests that the local moments of each chemical component (particularly Fe in Fe-Co, and both Fe and Ni in Fe-Ni) are well defined by the chemical composition of only a few close neighbor shells: up to the second nearest neighbor for Fe-Co, and up to the third one for Fe-Ni. In addition, the specific local arrangement of the solutes has a negligible effect on the local magnetic moments.

The models were also extended to predict the global magnetization of a system from the chemical short range orders and the global alloy concentration (the SRO-based model). In this way, they may be very useful to interpret experimental results. For instance, in both Fe-Co and Fe-Ni systems, experimental evolution of magnetization of random solid solutions as a function of the alloy concentration is well captured by the models, including the presence of a maximum, which occurs at a lower solute concentration in Fe-Ni than in Fe-Co. Furthermore, in the case of Fe-Co alloys, the larger magnetization of ordered (B2-like) structures compared with the random solutions observed experimentally, was successfully reproduced, and quantitatively explained by the enhanced local Fe moments in the ordered structures due to the higher average occupation of Co atoms at the first neighbor shells of Fe.

## Acknowledgments

This work was supported by the French-German ANR-DFG MAGIKID project. The DFT calculations were performed using the DARI-GENCI CPU resources within the A0050906020 project, and the CINECA-MARCONI supercomputer within the SISTEEL project.

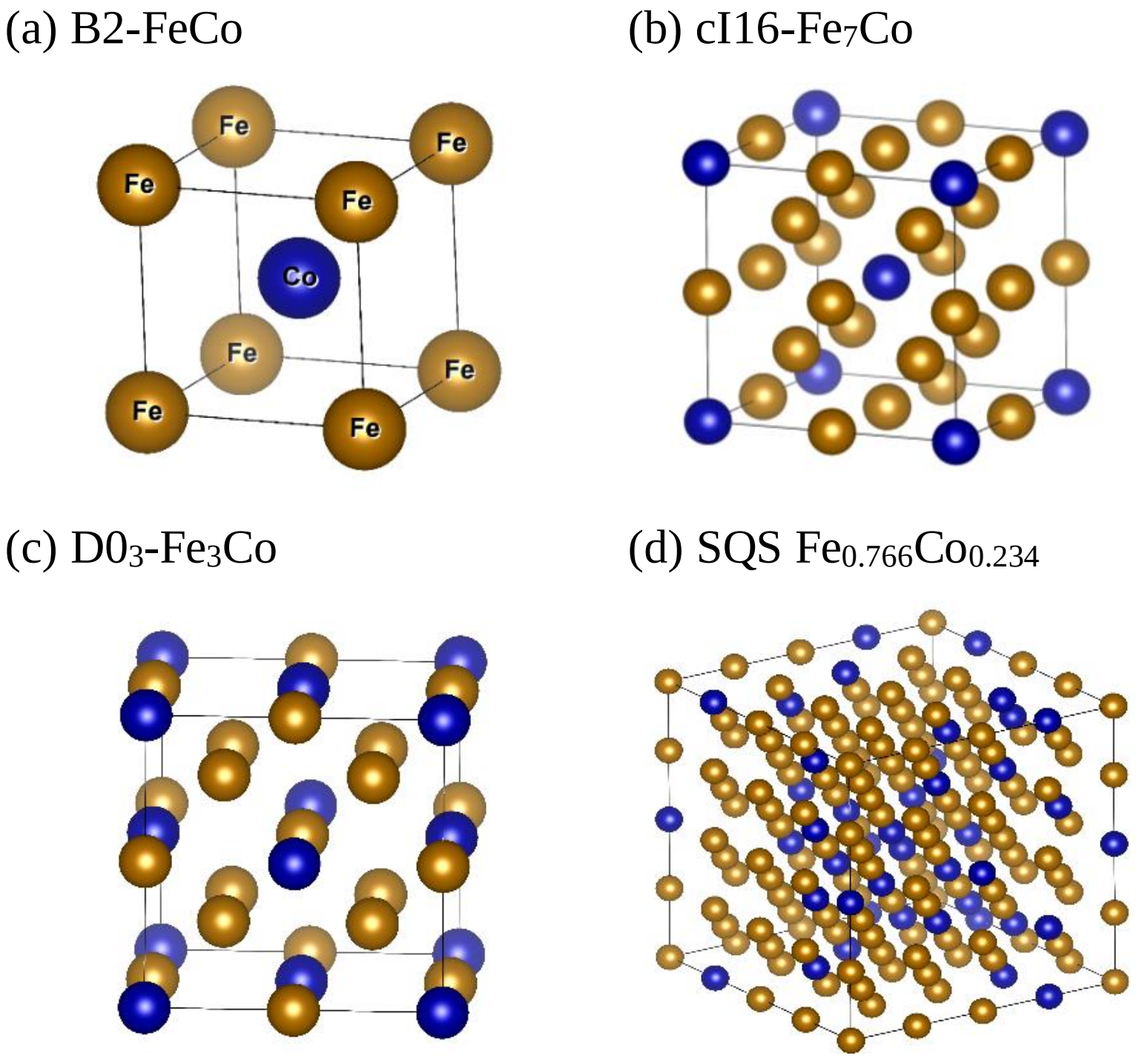

Fig. 1: Atomic configuration of some Fe-Co systems: (a) B2-FeCo, (b) cI16-$Fe_7Co$, (c) $D0_3$-$Fe_3Co$, and (d) a disordered (SQS) structure of $Fe_{0.766}Co_{0.234}$. In all the panels, Fe and Co atoms are represented by yellow and blue spheres, respectively.

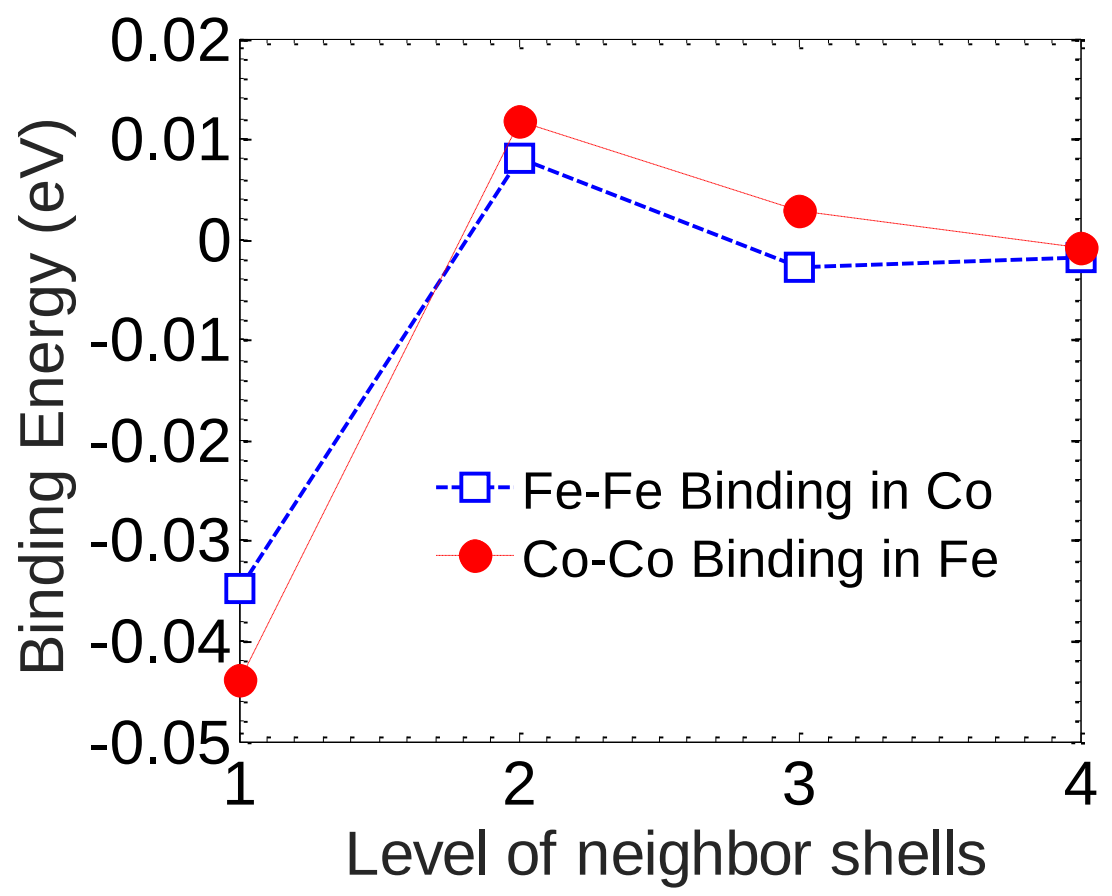

Fig. 2: Binding energies between two Fe (resp. two Co) atoms in a 128-site bcc Co (resp. Fe) supercell.

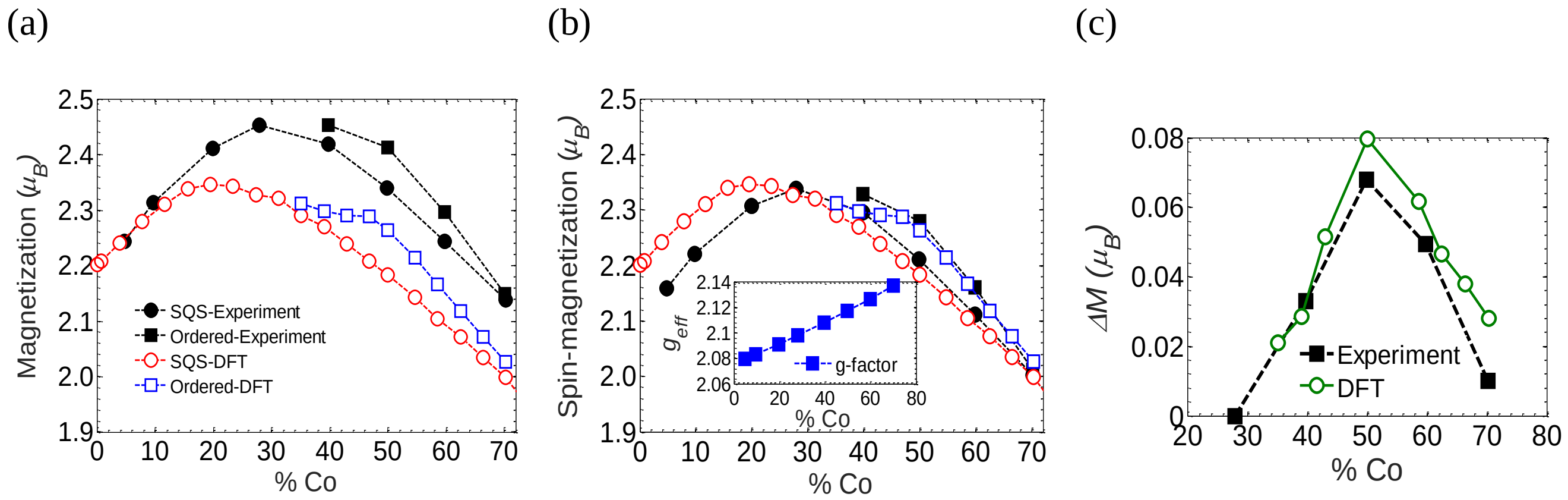

Fig 3: Average magnetic moments in ordered and SQS structures: (a) DFT results compared to the total experimental magnetization, (b) DFT results vs the spin magnetization estimated from experimental data. (c) The spin magnetization difference between the ordered and disordered structures obtained by experiment and DFT. The experimental values were extracted from Ref. 10.

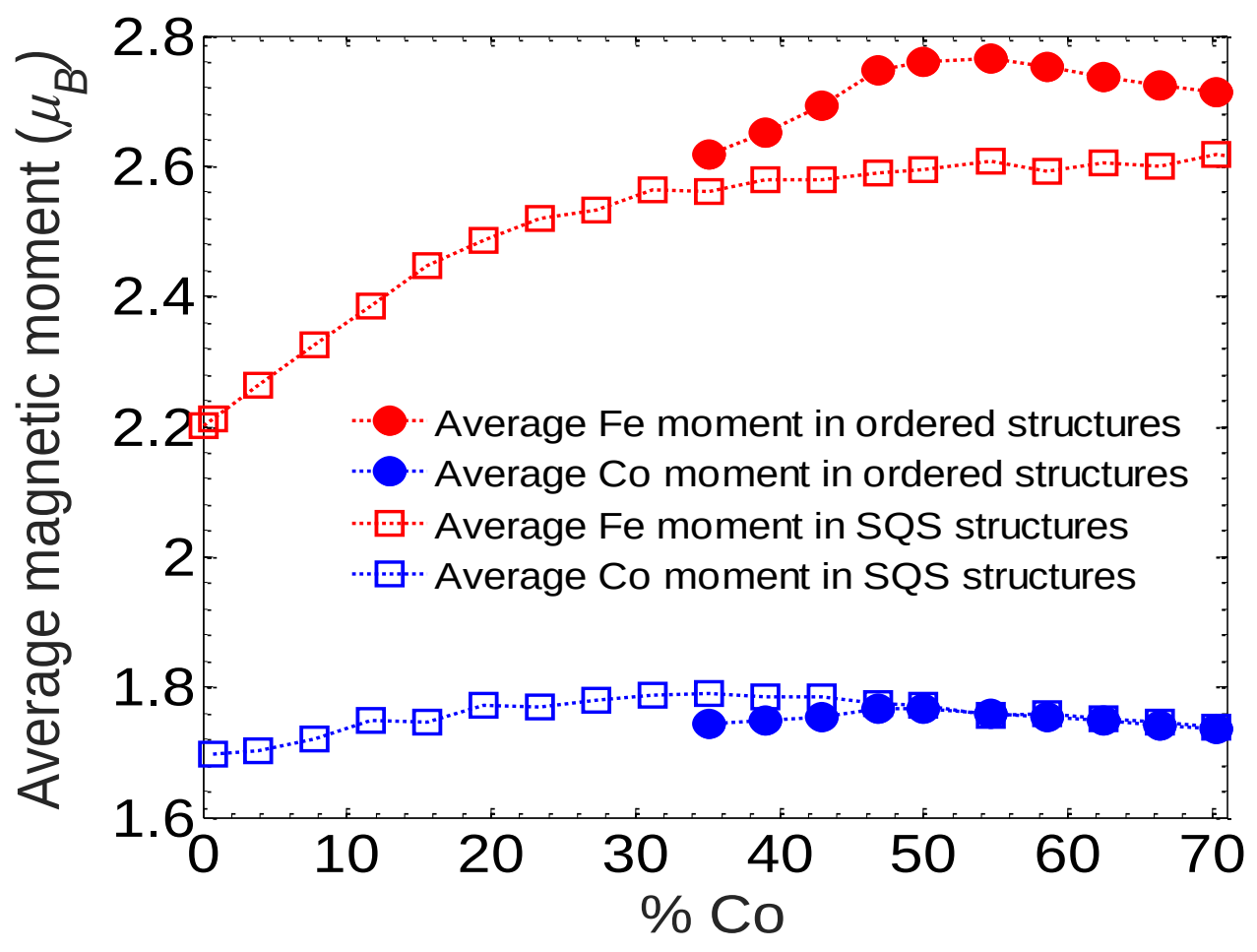

Fig 4: Average Fe and Co magnetic moments in SQSs and ordered structures of bcc Fe-Co systems.

(a)

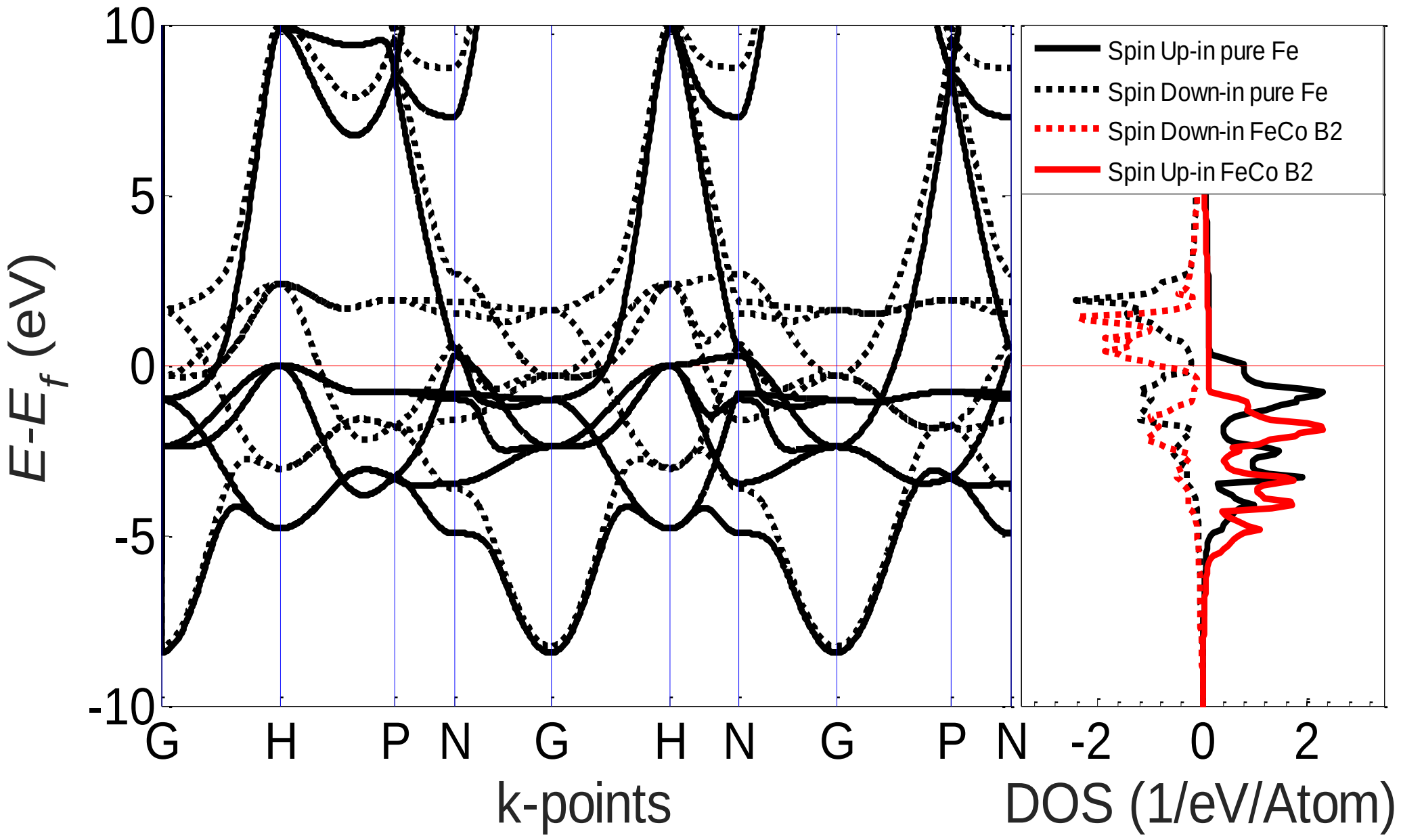

(b)

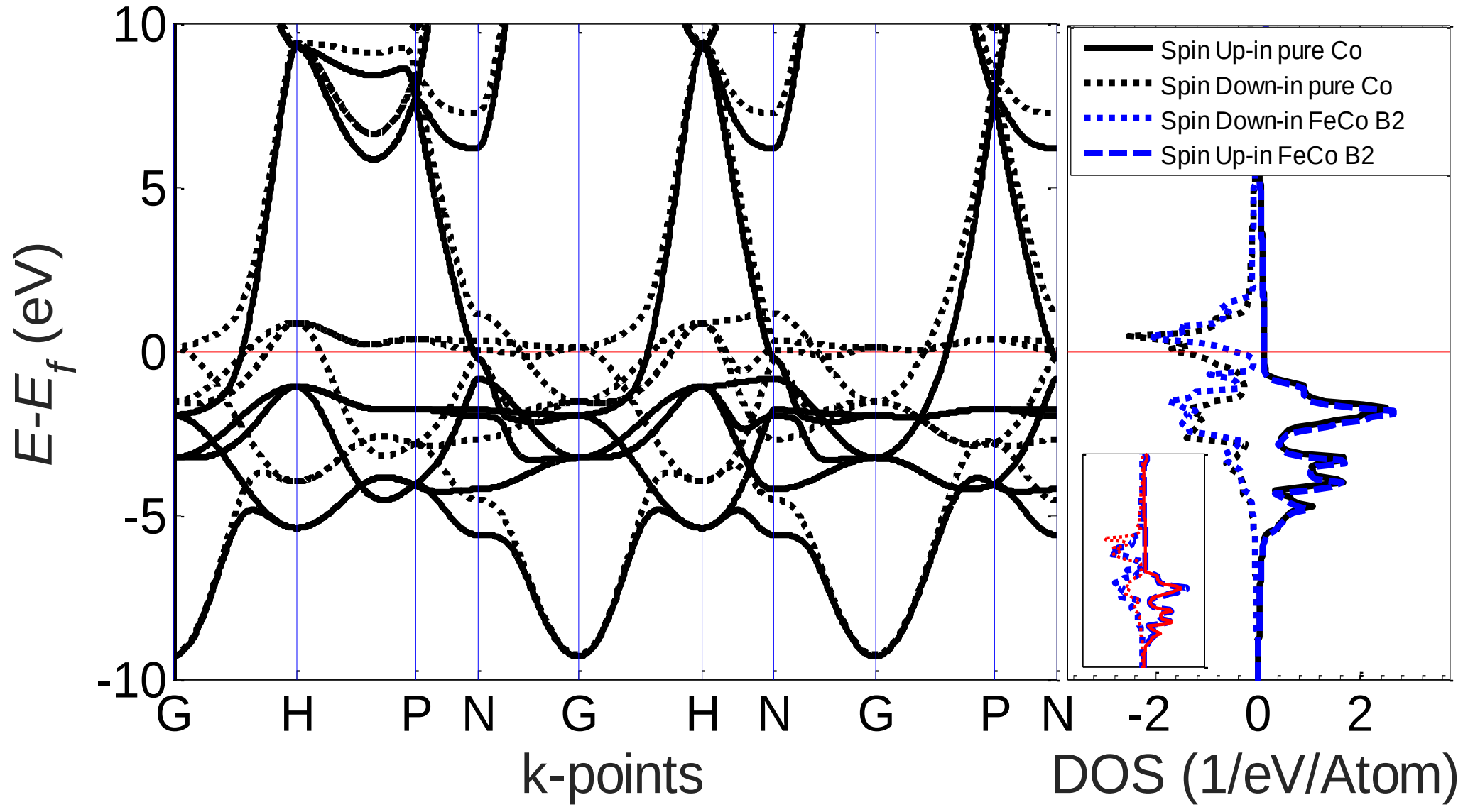

Fig 5: DFT electronic band structures and the DOSs of (a) Fe atoms in pure bcc Fe and in the Fe-Co B2, (b) Co in pure bcc Co and in the Fe-Co B2. All the structures are ferromagnetic.

(a)

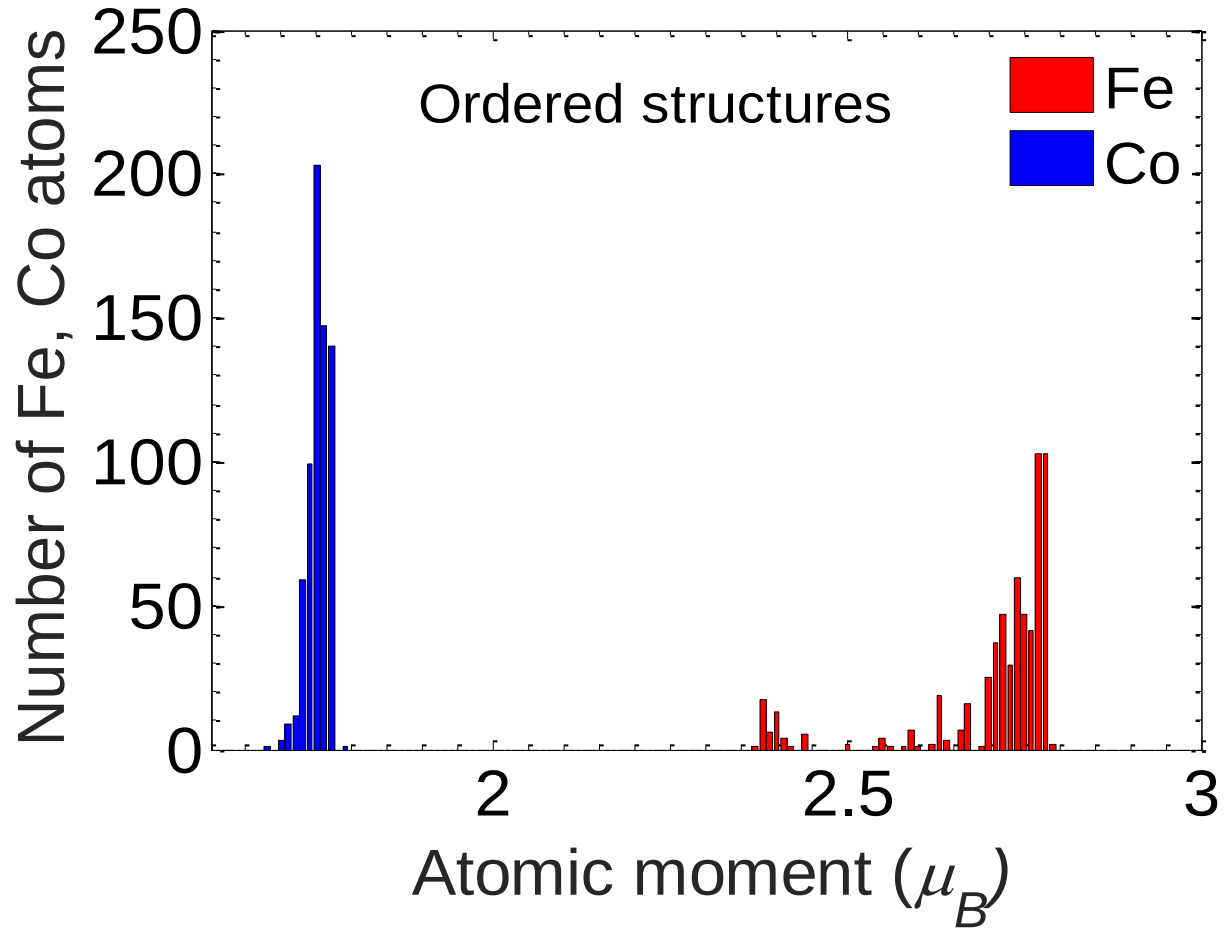

(b)

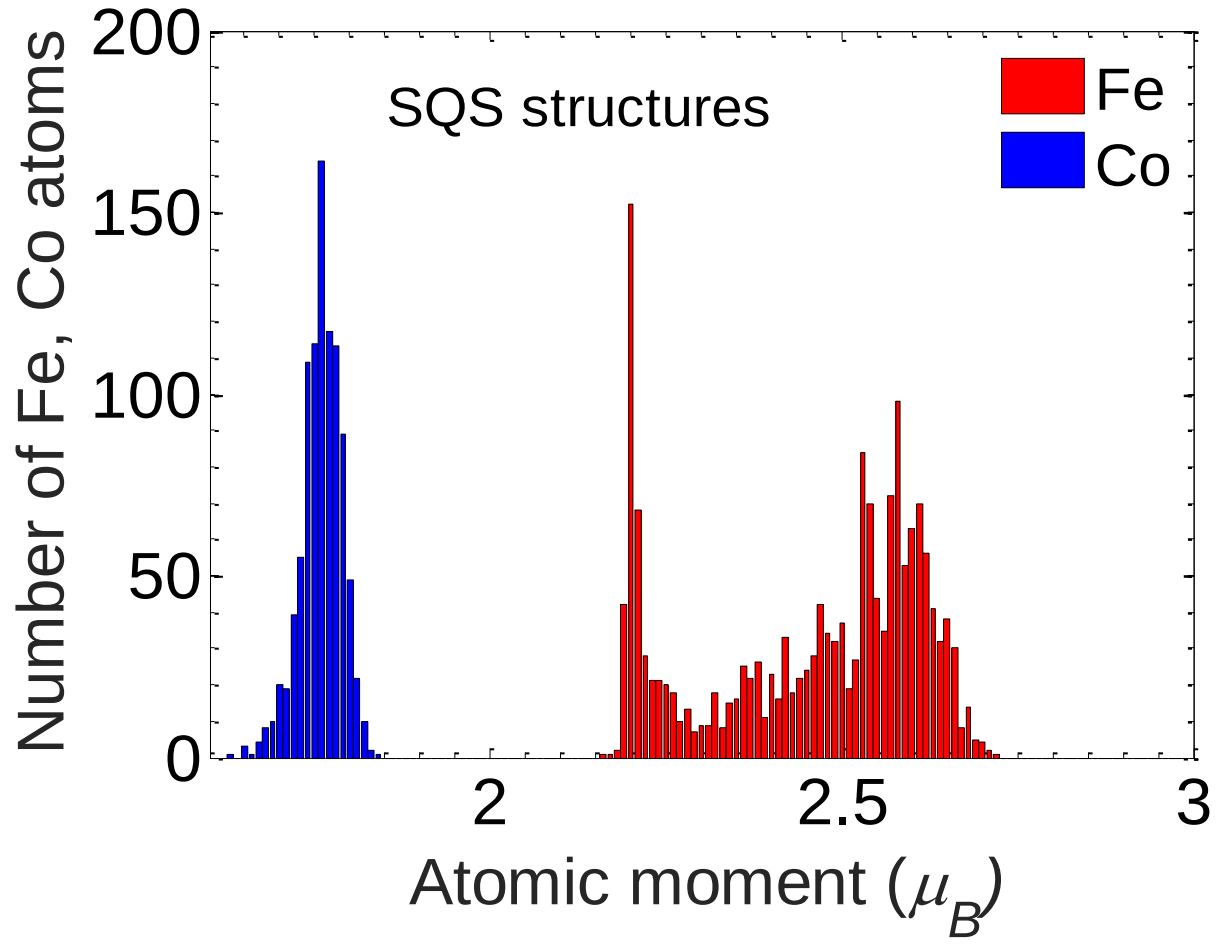

Fig 6: Distribution of atomic magnetic moments in (a) ordered and (b) disordered (SQS) Fe-Co structures.

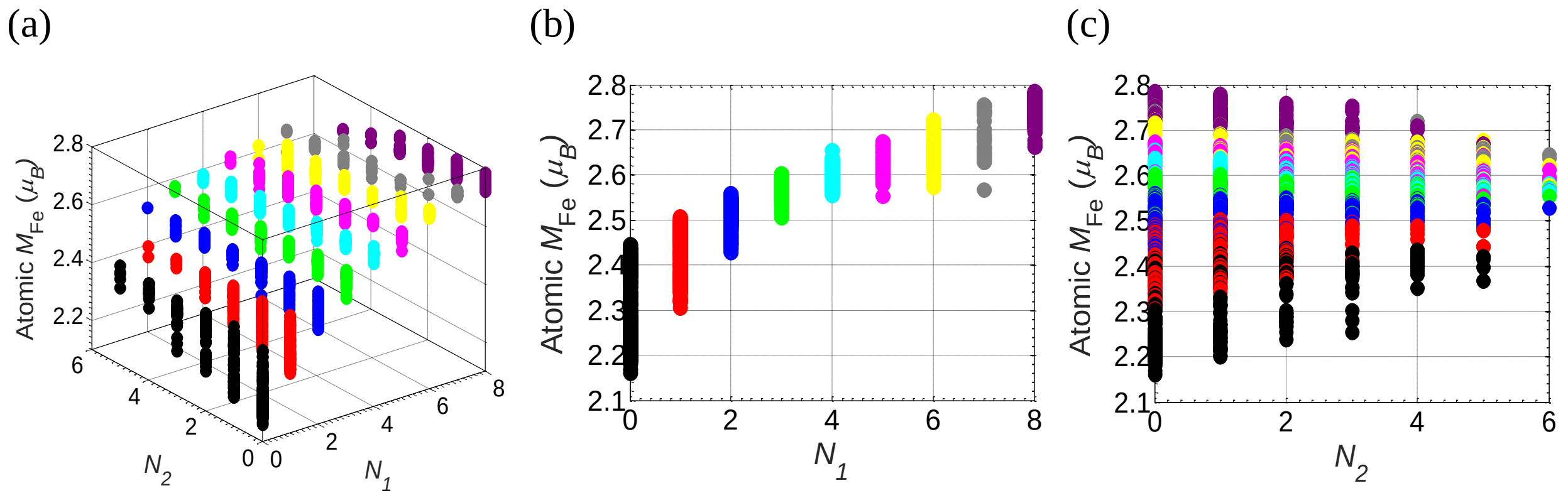

Fig 7: Local Fe magnetic moments ($M_{Fe}$) versus the number of Co atoms in the first ($N_1$) and second ($N_2$) neighbor shells of a Fe atom: (a) $M_{Fe}$ versus $N_1$ and $N_2$. (b) $M_{Fe}$ versus $N_1$ and (c) $M_{Fe}$ versus $N_2$. Data for different $N_1$ are represented by different colors. (DFT data)

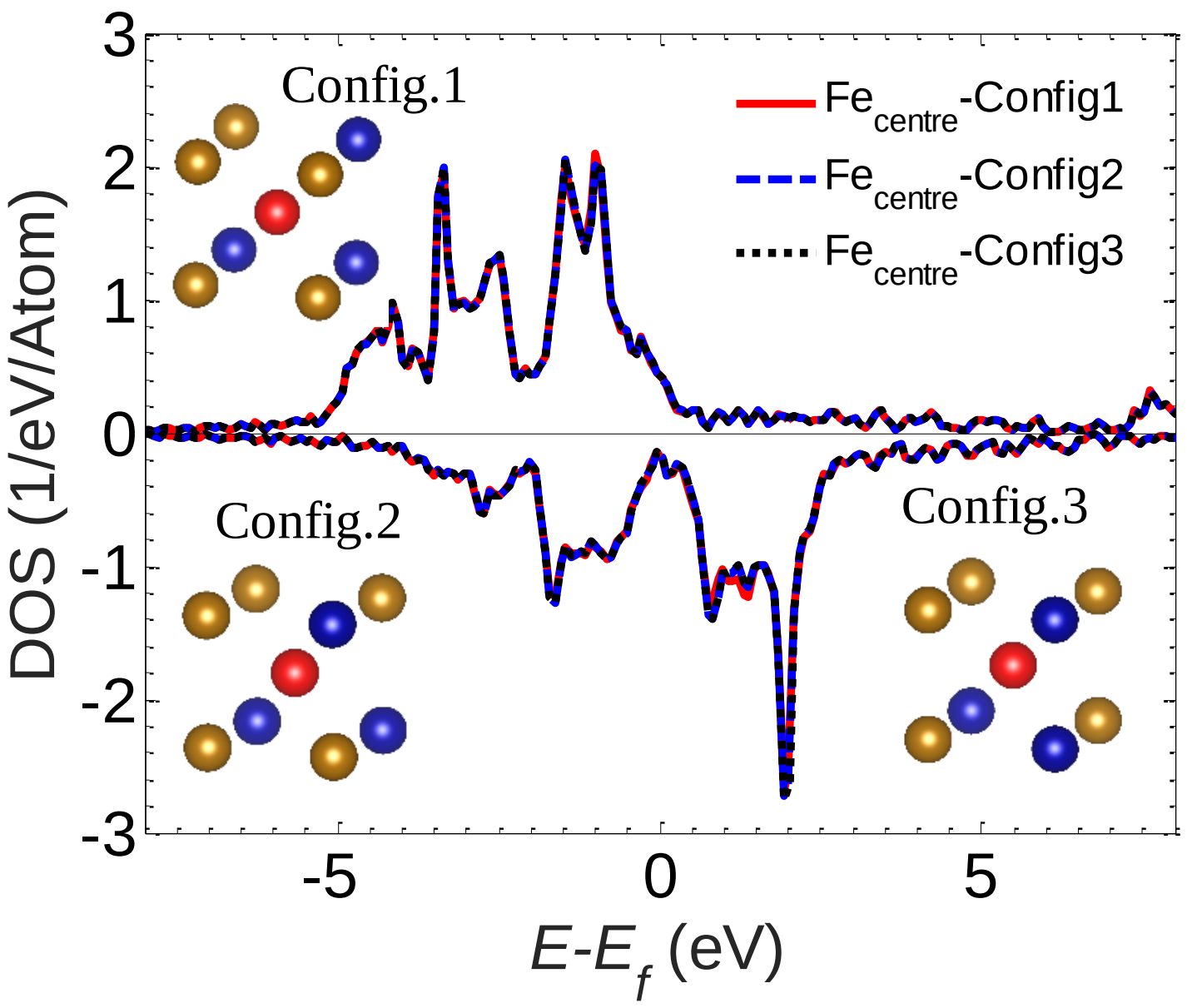

Fig 8: The DOS of the central Fe atom (marked as red) in three different configurations with distinct arrangements of Co atoms in its first shell. Only atoms in the first shell of the central Fe atoms are shown. Blue and dark-yellow spheres represent the Co and Fe atoms, respectively.

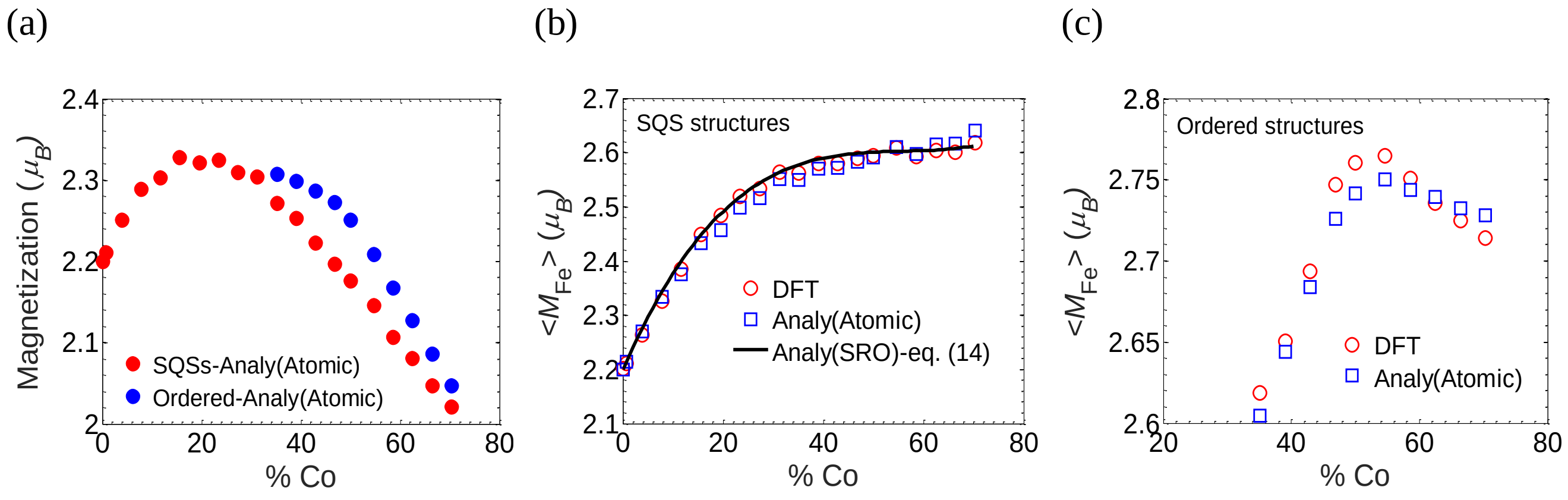

Fig 9: (a) The average magnetic moment obtained by the atomic model (Eq. 8) for both ordered and SQS structures. The magnetic moment of Fe produced by the analytical model in comparison with DFT data for (b) SQSs and (c) ordered Fe-Co structures. The results correspond to the 3rd order model.

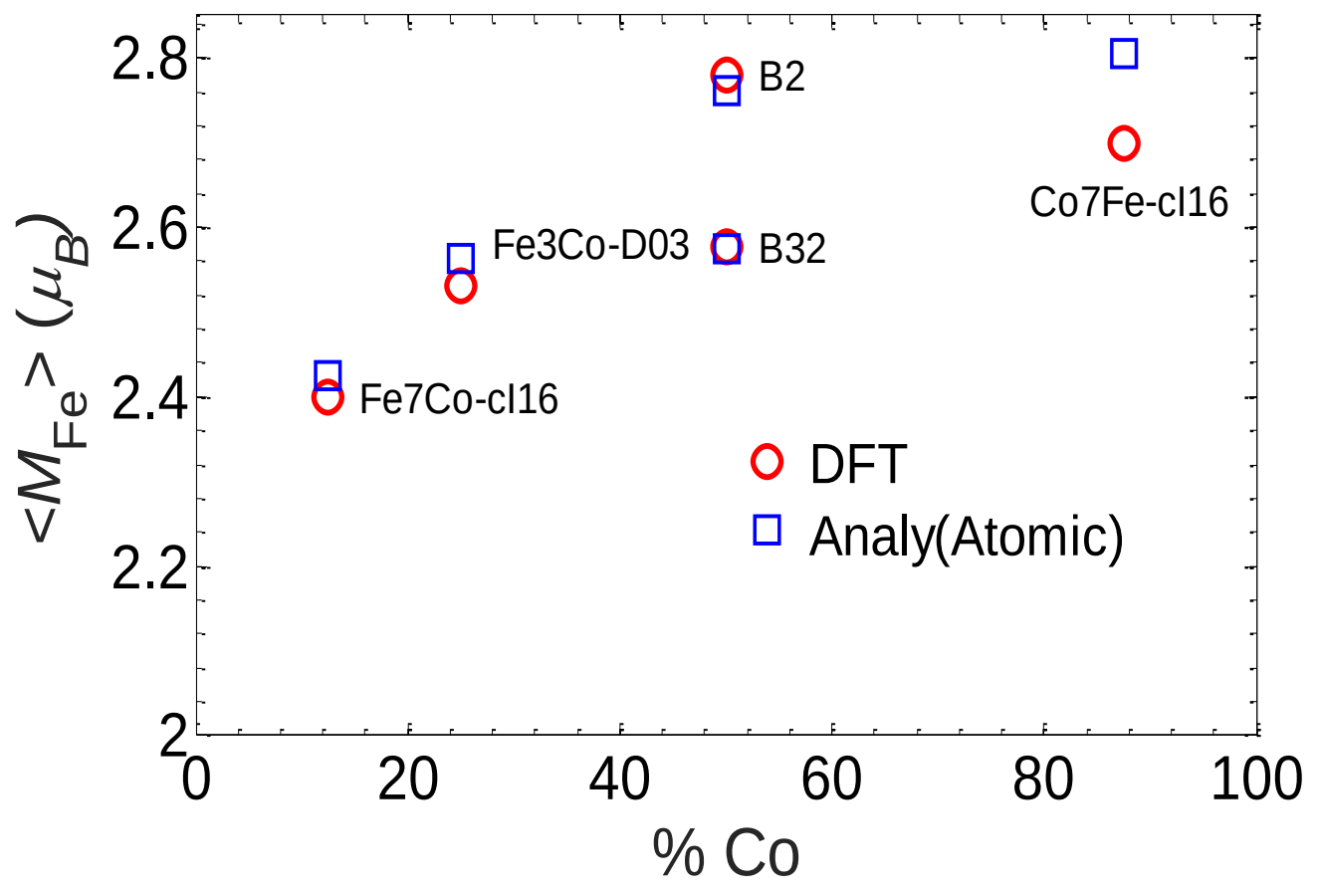

Fig 10: Average of Fe moments: Validation of the atomic model for some structures not included in the fitting data.

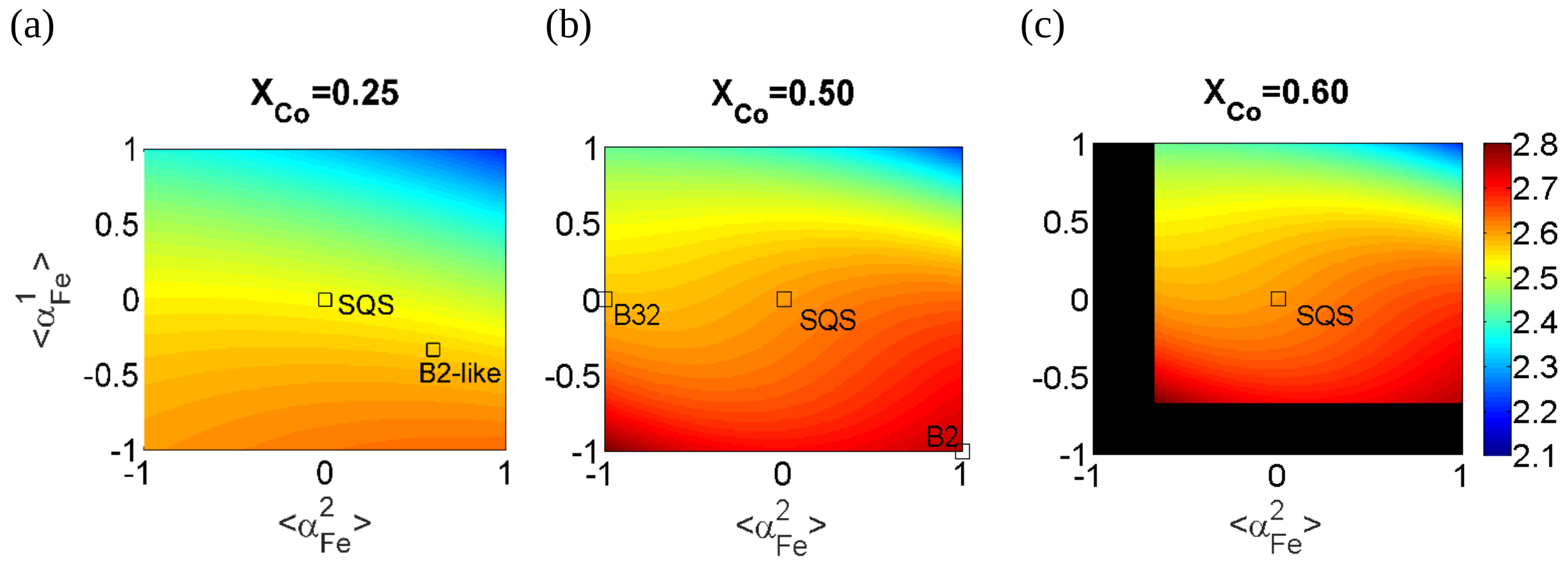

Fig. 11: Distribution of the Fe magnetic moment as a function of SRO parameters for various concentrations of Co: (a) 25% Co, (b) 50% Co and (c) 60% Co, using the SRO-based model. The black region in the panel (c) is a forbidden region.

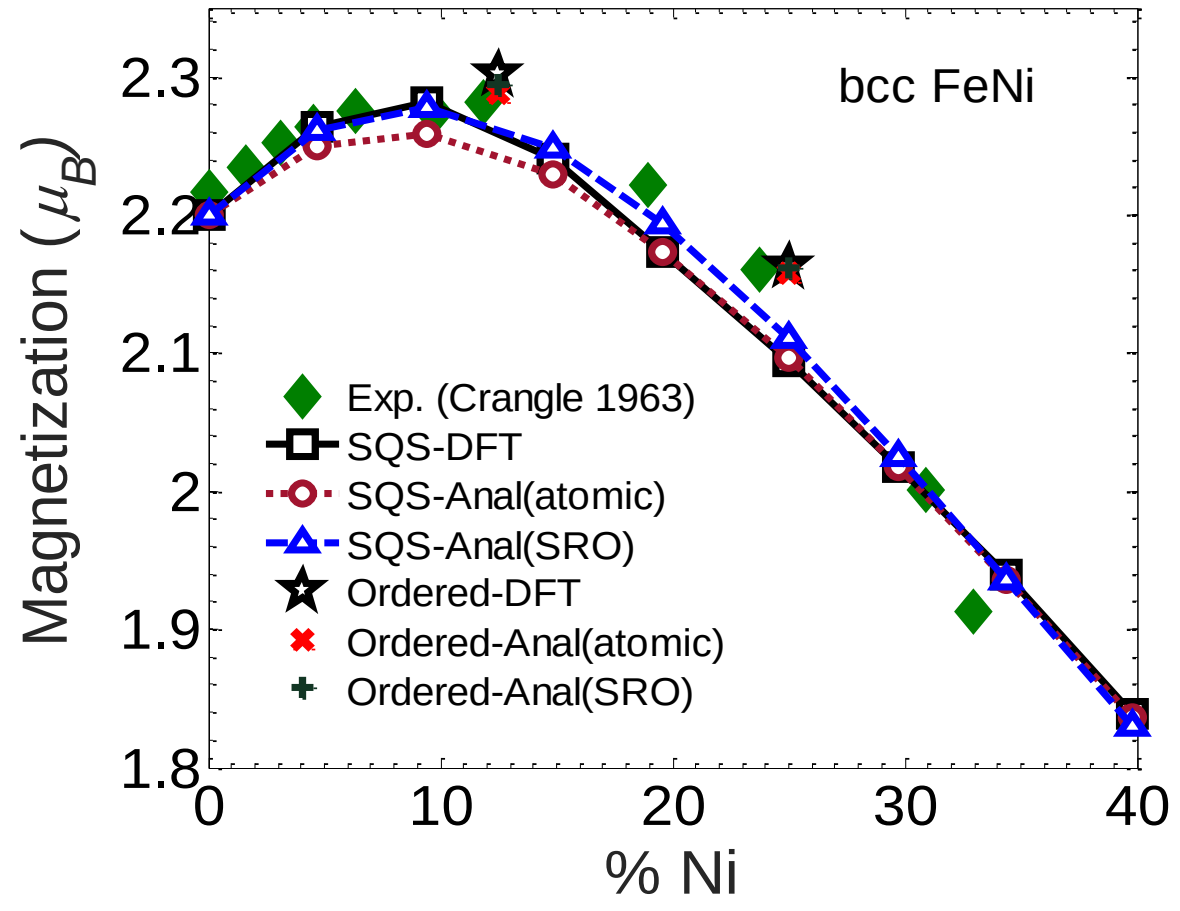

Fig. 12: Magnetization of bcc Fe-Ni obtained by the analytical models (both atomic and SRO-based) in comparison with DFT and experimental data.

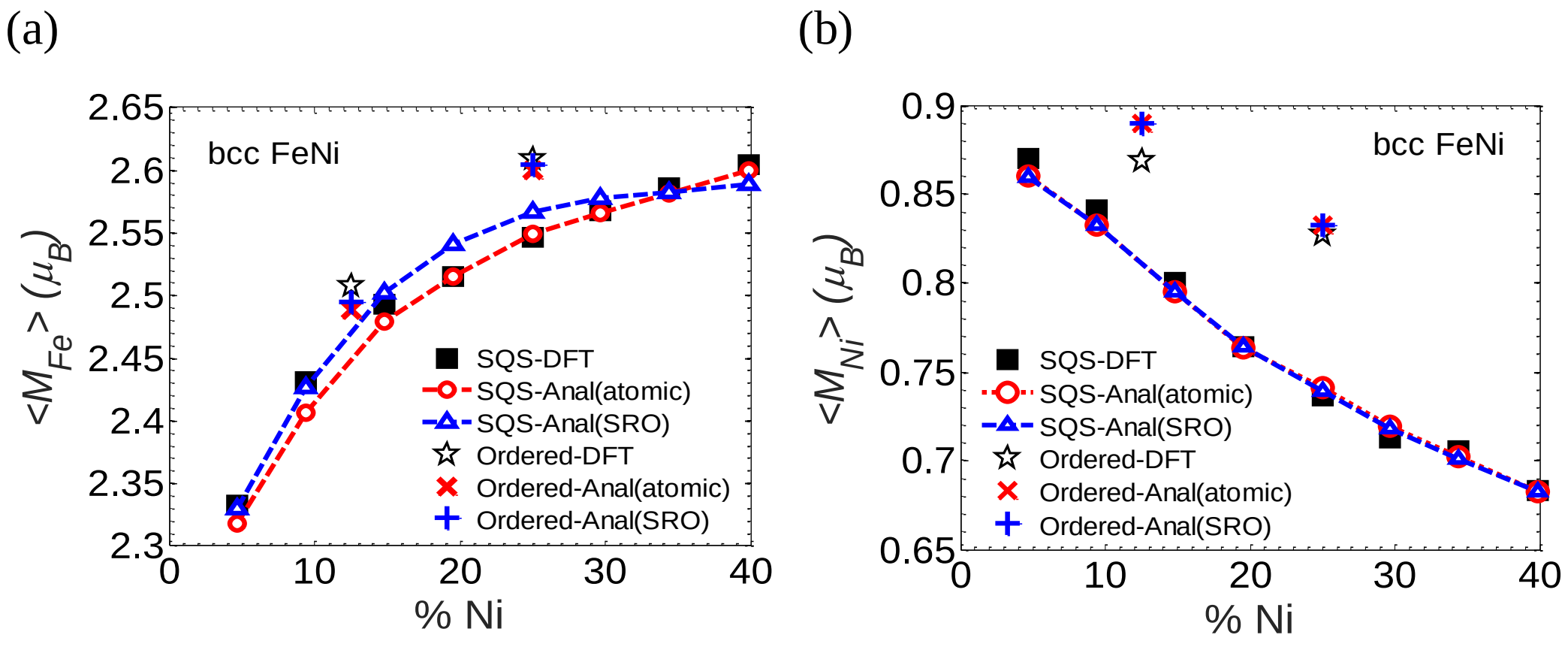

Fig. 13: Average moment of (a) Fe and (b) Ni versus Ni concentration, calculated by the analytical models and DFT.

Table I: Fitting parameters for the analytical model (Eq. 8) of Fe moments in Fe-Co systems

| | $P_0$ ($\mu_B$) | $A_1$ ($\mu_B$) | $A_2$ ($\mu_B$) | $A_3$ ($\mu_B$) | $B_1$ ($\mu_B$) | $B_2$ ($\mu_B$) | $B_3$ ($\mu_B$) | $C$ ($\mu_B$) | $D_1$ ($\mu_B$) | $D_2$ ($\mu_B$) |
|---|---|---|---|---|---|---|---|---|---|---|
| 2$^{nd}$ order | 2.2000 | 0.1365 | -0.0082 | _ | 0.0731 | -0.0062 | _ | -0.0115 | _ | _ |
| 3$^{rd}$ order | 2.2000 | 0.1910 | -0.0263 | 0.0014 | 0.0976 | -0.0148 | 0.0008 | -0.0318 | 0.0020 | 0.0020 |

Table II: Data of various FM structures: B2, cI16, B32, and $D0_3$

| Structures | SRO | $X_{Co}$ | DFT a (Å) | DFT $\langle M_{Fe}\rangle$ ($\mu_B$) | Analytical models $\langle M_{Fe}\rangle$ ($\mu_B$) Atomic Eq. (8) Fitting order of $N_1$, $N_2$: 2nd--3rd | Analytical models $\langle M_{Fe}\rangle$ ($\mu_B$) SRO-based Eqs. (10) + (13) Fitting order of $N_1$, $N_2$: 2nd--3rd |
|---|---|---|---|---|---|---|
| B2-FeCo | [-1, 1] | 0.5 | 2.840 | 2.779 | 2.767--2.762 | 2.767--2.762 |
| B32-FeCo | [0, -1] | 0.5 | 5.668 | 2.578 | 2.554--2.575 | 2.554--2.575 |
| cI16-$Fe_7Co$ | [-0.14286, -0.14286] | 0.125 | 5.681 | 2.399 | 2.389--2.426 | 2.392--2.432 |
| $D0_3$-$Fe_3Co$ | [-0.33333, -0.33333] | 0.25 | 5.688 | 2.532 | 2.548--2.564 | 2.566--2.571 |
| cI16-$Co_7Fe$ | [-0.14286 -0.14286] | 0.875 | 5.623 | 2.698 | 2.431--2.805 | 2.431--2.805 |

Table III: Fitting parameters for the models Eq. (17) for Fe and Ni moments in bcc FM Fe-Ni alloys up to three Ni neighbor shells using 3rd order polynomials:

| Parameter type | Parameters for Fe ($\mu_B$) | Parameters for Ni ($\mu_B$) |
|---|---:|---:|
| $P_0$ | 2.200000 | 0.890000 |
| $A_1$ | 0.201222 | -0.078890 |
| $A_2$ | 0.099028 | -0.005804 |
| $A_3$ | 0.081210 | -0.004098 |
| $B_{11}$ | -0.020708 | 0.004569 |
| $B_{12}$ | -0.037288 | 0.003376 |
| $B_{13}$ | -0.029148 | 0.000619 |
| $B_{22}$ | -0.018296 | -0.002222 |
| $B_{23}$ | -0.017682 | -0.001332 |
| $B_{33}$ | -0.008378 | 0.000277 |
| $C_{111}$ | 0.000462 | -0.000235 |
| $C_{112}$ | 0.003265 | -0.000372 |
| $C_{113}$ | 0.002050 | -0.000089 |
| $C_{122}$ | 0.002266 | 0.001011 |
| $C_{123}$ | 0.003190 | 0.000119 |
| $C_{133}$ | 0.001294 | 0.000319 |
| $C_{222}$ | 0.001190 | -0.000522 |
| $C_{223}$ | 0.001398 | 0.000764 |
| $C_{233}$ | 0.000547 | -0.000339 |
| $C_{333}$ | 0.000269 | -0.000028 |

Supplemental material for:

# Predicting magnetization of ferromagnetic binary Fe alloys from chemical short range order

**Van-Truong Tran, Chu-Chun Fu and Kangming Li**

DEN-Service de Recherches de Métallurgie Physique, CEA, Université Paris-Saclay, F-91191 Gif-sur-Yvette, France

Table S1: Data for Fe-Co alloys with three different concentrations and with various SROs.

| Structures Fe-Co 128 atoms | SRO | $X_{Co}$ | DFT $\langle M_{Fe} \rangle$ ($\mu_B$) | Analytical models $\langle M_{Fe} \rangle$ ($\mu_B$) Atomic Eq. (8) Fitting order of $N_1$, $N_2$: 2nd--3rd | $\langle M_{Fe} \rangle$ ($\mu_B$) SRO-based Eqs. (10) + (13) 2nd--3rd |
|---|---|---|---|---|---|
| NCo=30 | [-0.18639, 0.20181] | 0.2344 | 2.525 | 2.486--2.506 | 2.509--2.540 |
| | [-0.24082, 0.14376] | | 2.535 | 2.503--2.525 | 2.520--2.548 |
| | [-0.30612, 0.59365] | | 2.533 | 2.464--2.479 | 2.509--2.546 |
| | [0.11837, 0.10023] | | 2.495 | 2.433--2.448 | 2.462--2.497 |
| | [0.18367, 0.17279] | | 2.484 | 2.418--2.437 | 2.446--2.484 |
| NCo=64 | [-0.14844, 0.14583] | 0.5000 | 2.616 | 2.641--2.619 | 2.665--2.619 |
| | [-0.30469, 0.12500] | | 2.637 | 2.665--2.635 | 2.681--2.627 |
| | [-0.00781, 0.00000] | | 2.599 | 2.625--2.598 | 2.641--2.601 |
| | [-0.49219, 0.51042] | | 2.671 | 2.689--2.671 | 2.716--2.666 |
| | [0.10156, 0.12500] | | 2.585 | 2.590--2.577 | 2.625--2.598 |
| NCo=90 | [-0.13216, 0.13918] | 0.7031 | 2.640 | 2.636--2.656 | 2.654--2.636 |
| | [-0.20702, 0.20156] | | 2.654 | 2.646--2.670 | 2.662--2.651 |
| | [0.00819, 0.00195] | | 2.608 | 2.612--2.630 | 2.634--2.611 |
| | [-0.42222, 0.50097] | | 2.705 | 2.690--2.725 | 2.700--2.714 |
| | [0.12982, 0.12671] | | 2.590 | 2.630--2.609 | 2.649--2.604 |

Table S2: Results obtained by the analytical models (both atomic and SRO-based) and DFT for some bcc Fe-Ni structures with different SROs.

| $X_{Ni}$ | SRO | Magnetization ($\mu_B$) | | | <$M_{Fe}$> ($\mu_B$) | | | <$M_{Ni}$> ($\mu_B$) | | |
|---|---|---|---|---|---|---|---|---|---|---|
| | | | *Analytical model* | | | *Analytical model* | | | *Analytical model* | |
| | | DFT | *Atomic* | *SRO* | DFT | *Atomic* | *SRO* | DFT | *Atomic* | *SRO* |
| 0.1562 | [-0.1852, 0.5654, 0.3086] | 2.227 | 2.192 | 2.242 | 2.494 | 2.453 | 2.511 | 0.787 | 0.784 | 0.788 |
| 0.2344 | [-0.3061, 0.5646, 0.3687] | 2.149 | 2.133 | 2.273 | 2.571 | 2.553 | 2.603 | 0.773 | 0.758 | 0.767 |
| 0.2344 | [-0.2408, 0.1438, 0.1873] | 2.156 | 2.160 | 2.175 | 2.573 | 2.576 | 2.592 | 0.794 | 0.801 | 0.814 |
| 0.2344 | [-0.3061, 0.5937, 0.4050] | 2.142 | 2.122 | 2.171 | 2.562 | 2.542 | 2.603 | 0.770 | 0.751 | 0.758 |
| 0.3125 | [-0.4546, 0.6242, 0.5091] | 2.041 | 2.022 | 2.060 | 2.628 | 2.616 | 2.668 | 0.749 | 0.716 | 0.723 |
| 0.3906 | [-0.2472, 0.1357, 0.1904] | 1.880 | 1.874 | 1.889 | 2.634 | 2.626 | 2.634 | 0.705 | 0.702 | 0.727 |